\documentclass[letterpaper]{article}
\pdfoutput=1
\usepackage{aaai16}

\usepackage{comment}
\includecomment{report}
\excludecomment{paper}

\usepackage{amsmath}
\usepackage{graphicx}
\usepackage[utf8]{inputenc}
\usepackage[T1]{fontenc}
\usepackage{times}
\usepackage{txfonts} %
\usepackage[scaled=0.9]{helvet} %
\usepackage{color}
\usepackage{import}

\newcommand{\ourtitle}{Column-Oriented Datalog Materialization for Large Knowledge Graphs}

\usepackage{url}
\usepackage[bookmarks=false,pdfpagelabels,bookmarksnumbered,bookmarksopen,bookmarksopenlevel=1,colorlinks=true,
citecolor=black, filecolor=black, linkcolor=black, menucolor=black,
urlcolor=black, pdftitle={\ourtitle}, pdfauthor={Jacopo Urbani, Ceriel Jacobs,
    Markus Kr\"{o}tzsch},
pdfkeywords={Datalog, column store, semi-naive evaluation, in-memory computation, inferencing, reasoning, knowledge graphs}]{hyperref} 

\begin{report}
\makeatletter
\def\copyright@text{This is the extended version of the eponymous paper published at the AAAI 2016 \cite{UJK2016:columnDatalog}.}
\makeatother
\end{report}

\makeatletter{}%
\usepackage{xspace}

\newcommand{\tableTopSpace}{\rule{0mm}{2ex}} %

\newtheorem{lemma}{Lemma}
\newtheorem{theorem}[lemma]{Theorem}

\newcommand{\qed}{\mbox{}\hfill$\square$}

\newcommand{\tuple}[1]{\langle{#1}\rangle}
\newcommand{\defeq}{\coloneqq} %
\newcommand{\arity}{\text{\sf{ar}}} %
\renewcommand{\vec}[1]{\boldsymbol{#1}}

\newcommand{\Inter}{\mathcal{I}} %

\newcommand{\lang}[1]{\ensuremath{\mathbf{#1}}} %
\newcommand{\Clang}{\lang{C}\xspace} %
\newcommand{\Plang}{\lang{P}\xspace} %
\newcommand{\Vlang}{\lang{V}\xspace} %

\newcommand{\aprogram}{\mathbb{P}} %
\newcommand{\arule}{r} %

\newcommand{\steprule}[1]{\text{\sf{rule}}[#1]} %

\newcommand{\fakeparagraph}[1]{\vspace{1mm}\noindent\textbf{#1}~~}
\newcommand{\toolname}{VLog\xspace}
\begin{paper}
\title{\ourtitle}
\end{paper}
\begin{report}
\title{\ourtitle\\(Extended Technical Report)}
\end{report}

\author{Jacopo Urbani\\
    Dept. Computer Science\\
 VU University Amsterdam\\
 Amsterdam, The Netherlands\\
 jacopo@cs.vu.nl
 \And
Ceriel Jacobs\\
    Dept. Computer Science\\
 VU University Amsterdam\\
 Amsterdam, The Netherlands\\
 c.j.h.jacobs@vu.nl
 \And
 Markus Kr\"{o}tzsch\\
 Faculty of Computer Science\\
 Technische Universit\"{a}t Dresden\\
 Dresden, Germany\\
 markus.kroetzsch@tu-dresden.de
}

\begin{document}

\maketitle

\makeatletter{}%
\begin{abstract}
The evaluation of Datalog rules over large Knowledge Graphs (KGs) is essential for many
applications.
In this paper, we present a new method of materializing Datalog inferences, which combines a column-based memory layout with novel optimization methods that avoid redundant inferences at runtime.
The pro-active caching of certain subqueries further increases efficiency.
Our empirical evaluation shows that this approach can often match or even surpass the performance of state-of-the-art systems, especially under restricted resources.
\end{abstract} %

\makeatletter{}%
\section{Introduction} %

Knowledge graphs (KGs) are widely used in industry and academia to represent large collections of structured knowledge. While many types of graphs are in use, they all rely on simple,
highly-normalized data models that can be used to uniformly represent information from many diverse sources. On the Web, the most prominent such format is RDF \cite{rdf11-concepts},
and large KGs such as Bio2RDF \cite{bio2rdf13}, DBpedia \cite{dbpedia09}, Wikidata \cite{VK:wikidata14}, and YAGO \cite{HSBW12:yago2} are published in this format.

The great potential in KGs is their ability to make connections -- in a literal sense --
between heterogeneous and often incomplete data sources.
Inferring implicit information from KGs is therefore essential in many applications, such as ontological reasoning, data integration, and information extraction.
The rule-based language Datalog offers a common foundation for specifying such inferences
\cite{Alice}.
While Datalog rules are rather simple types of \emph{if-then} rules, their recursive
nature is making them powerful. Many inference tasks can be captured in this
framework, including many types of ontological reasoning commonly used with RDF.
Datalog thus provides an excellent basis for exploiting KGs to the full.

Unfortunately, the implementation of Datalog inferencing on large KGs remains
very challenging. The task is worst-case time-polynomial in the size
of the KG, and hence tractable in principle, but huge
KGs are difficult to manage. A preferred approach is therefore to \emph{materialize} (i.e., pre-compute) inferences.
Modern DBMS such as Oracle~11g and OWLIM materialize KGs of 100M--1B edges
in times ranging from half an hour to several days \cite{KWE10:oracleowlrl,B+11:owlimswj}.
Research prototypes such as Marvin \cite{OKASTH:Marvin2009}, C/MPI \cite{WH09:parallelRDFS}, WebPIE \cite{webpie}, and DynamiTE \cite{dynamite} achieve scalability by using
parallel or distributed computing, but often require significant hardware resources.
\citeauthor{webpie}, e.g., used up to 64 high-end machines to materialize a KG with 100B edges
in 14 hours \shortcite{webpie}. In addition, all the above systems only support (fragments of) the OWL~RL ontology language, which is subsumed by Datalog but significantly simpler.

\citeauthor{rdfox} have recently presented a completely new approach to this problem \shortcite{rdfox}.
Their system RDFox exploits fast main-memory computation and parallel processing.
A groundbreaking insight of this work is that this approach allows processing mid-sized KGs on commodity machines. This has opened up a new research field
for in-memory Datalog systems, and \citeauthor{MNPH15:rdfoxIncrementalEquality} have presented
several advancements
\shortcite{MNPH15:rdfoxIncrementalEquality,MNPH15:rdfoxEqualityRewriting,MNPH15:rdfoxDRedUpdate}.

Inspired by this line of research, we present a new approach
to in-memory Datalog materialization. Our goal is to further reduce memory consumption to
enable even larger KGs to be processed on even simpler computers.
To do so, we propose to maintain inferences in an ad-hoc column-based storage layout.
In contrast to traditional row-based layouts,
where a data table is represented as a list of tuples (rows), column-based approaches
use a tuple of columns (value lists) instead. This
enables more efficient joins~\cite{IGNMMK12:MonetDBColumnResearch} and
effective, yet simple data compression schemes~\cite{abadi_integrating_2006}.
However, these advantages are set off by the comparatively high cost of updating
column-based data structures \cite{AMMH09:SWstore}.
This is a key challenge for using this technology
during Datalog materialization, where frequent insertions of large
numbers of newly derived inferences need to be processed.
Indeed, to the best of our knowledge, no materialization approach has yet
made use of columnar data structures.
Our main contributions are as follows:

\begin{itemize}
\item We design novel column-based data structures for in-memory Datalog materialization.
Our memory-efficient design organizes inferences by rule and inference step.
\item We develop novel optimization techniques that reduce the amount of data that is considered during materialization.
\item We introduce a new \emph{memoization} method~\cite{ai2003} that caches results of selected subqueries proactively, improving the performance of our procedure and optimizations.
\item We evaluate a prototype implementation or our approach.
\end{itemize}

Evaluation results show that our approach can significantly reduce the amount of
main memory needed for materialization, while maintaining competitive
runtimes. This allowed us to materialize fairly large graphs on commodity hardware. Evaluations also show that our optimizations contribute significantly
to this result.

\begin{paper}
Proofs for the claims in this paper can be found in an extended technical report \cite{techreport}.
\end{paper}
\begin{report}
Proofs for the claims in this paper can be found in the appendix.
\end{report}

\makeatletter{}%
\section{Preliminaries}\label{sec:prelims}

We define Datalog in the usual way; details can be found in the textbook by \citeauthor{Alice} \shortcite{Alice}. We assume a fixed signature consisting of an infinite set \Clang of \emph{constant symbols}, an infinite set \Plang of \emph{predicate symbols}, and an infinite set \Vlang of \emph{variable symbols}. Each predicate $p\in\Plang$ is associated with an \emph{arity} $\arity(p)\geq 0$.
A \emph{term} is a variable $x\in\Vlang$ or a constant $c\in\Clang$. We use symbols $s$, $t$ for terms; $x$, $y$, $z$, $v$, $w$ for variables; and $a$, $b$, $c$ for constants. Expressions like $\vec{t}$, $\vec{x}$, and $\vec{a}$ denote finite lists of such entities.
An \emph{atom} is an expression $p(\vec{t})$ with $p\in\Plang$ and $|\vec{t}|=\arity(p)$. A \emph{fact} is a variable-free atom. A \emph{database instance} is a finite set $\Inter$ of facts.
A \emph{rule} $\arule$ is an expression of the form
\begin{equation}
H \leftarrow B_1,\ldots,B_n \label{eq:rule}
\end{equation}
where $H$ and $B_1$, \ldots, $B_n$ are \emph{head} and \emph{body} atoms, respectively. We assume rules to be \emph{safe}: every variable in $H$ must also occur in some $B_i$. A \emph{program} is a finite set $\aprogram$ of rules.

Predicates that occur in the head of a rule are called \emph{intensional (IDB) predicates}; all other predicates are \emph{extensional (EDB)}. IDB predicates must not appear in databases. Rules with at most one IDB predicate in their body are \emph{linear}.%

A \emph{substitution} $\sigma$ is a partial mapping $\Vlang\to\Clang\cup\Vlang$. Its application to atoms and rules is defined as usual. For a set of facts $\Inter$ and a rule $\arule$ as in \eqref{eq:rule}, we define $\arule(\Inter)\defeq\{H\sigma\mid H\sigma\text{ is a fact, and } B_i\sigma\in\Inter\text{ for all }1\leq i\leq n\}$.
For a program $\aprogram$, we define $\aprogram(\Inter)\defeq\bigcup_{\arule\in\aprogram} \arule(\Inter)$, and shortcuts
$\aprogram^0(\Inter)\defeq \Inter$ and $\aprogram^{i+1}(\Inter)\defeq \aprogram(\aprogram^i(\Inter))$. The set $\aprogram^\infty(\Inter)\defeq \bigcup_{i\geq 0}\aprogram^i(\Inter)$ is the \emph{materialization of $\Inter$ with $\aprogram$}. This materialization is finite, and contains all facts that are logical consequences of $\Inter\cup\aprogram$.

Knowledge graphs are often encoded in the RDF data model~\cite{rdf11-concepts}, which represents labelled graphs as sets of triples of the form
$\tuple{\textsf{subject},\textsf{property},\textsf{object}}$. Technical details are not relevant here.
Schema information for RDF graphs can be expressed  using the W3C OWL Web Ontology Language.
Since OWL reasoning is complex in general, the standard
offers three lightweight profiles that simplify this task. In particular, OWL reasoning can be captured with Datalog in all three cases, as shown by
\citeauthor{Kroetzsch11:elreason} \shortcite{Kroetzsch11:elreason,Kroetzsch12:owlrl} and (implicitly by translation to path queries) by \citeauthor{BKPR14:nobda} \shortcite{BKPR14:nobda}.

\newcommand{\edbTriple}{\textsf{triple}}
\newcommand{\idbTriple}{\textsf{T}}
\newcommand{\idbInv}{\textsf{Inverse}}
\newcommand{\constOwlInvOf}{\textsf{owl:inverseOf}}
\newcommand{\constTransRel}{\textsf{hasPart}}
\newcommand{\constTransRelInv}{\textsf{partOf}}
\newcommand{\idbTransRelDomain}{\textsf{Compound}}%
The simplest encoding of RDF data for Datalog is to use a ternary EDB predicate $\edbTriple$ to represent triples. We use a simple Datalog program
as a running example:
{%
\begin{align}
\idbTriple(x,v,y) \leftarrow{}
    & \edbTriple(x,v,y)
\label{eq:ruleedbidb}
\\\displaybreak[0]
\idbInv(v,w) \leftarrow{}
    & \idbTriple(v,\textsf{\constOwlInvOf},w)
\label{eq:ruleinv}
\\\displaybreak[0]
\idbTriple(y,w,x) \leftarrow{}
    & \idbInv(v,w),\idbTriple(x,v,y)
\label{eq:ruleinvone}
\\\displaybreak[0]
\idbTriple(y,v,x) \leftarrow{}
    & \idbInv(v,w),\idbTriple(x,w,y)
\label{eq:ruleinvtwo}
\\\displaybreak[0]
\idbTriple(x,\constTransRel,z) \leftarrow{}
    & \idbTriple(x,\constTransRel,y),\idbTriple(y,\constTransRel,z)
\label{eq:ruletrans}
\end{align}}%
To infer new triples, we need an IDB predicate
$\idbTriple$, initialised in rule \eqref{eq:ruleedbidb}. Rule \eqref{eq:ruleinv}
``extracts'' an RDF-encoded OWL statement
that declares a property to be the inverse of another.
Rules \eqref{eq:ruleinvone} and \eqref{eq:ruleinvtwo}
apply this information to derive inverted triples.
Finally, rule \eqref{eq:ruletrans} is a typical transitivity rule for the RDF property $\constTransRel$.

\newcommand{\constAbbOwlInvOf}{\textsf{iO}}
\newcommand{\constAbbTransRel}{\textsf{hP}}
\newcommand{\constAbbTransRelInv}{\textsf{pO}}
\newcommand{\constA}{\textsf{a}}
\newcommand{\constB}{\textsf{b}}
\newcommand{\constC}{\textsf{c}}
We abbreviate $\constTransRel$, $\constTransRelInv$ and $\constOwlInvOf$ by
$\constAbbTransRel$, $\constAbbTransRelInv$ and $\constAbbOwlInvOf$, respectively.
Now consider a database $\Inter=
\{\edbTriple(\constA,\allowbreak\constAbbTransRel, \constB),\allowbreak
\edbTriple(\constB,\allowbreak\constAbbTransRel, \constC),\allowbreak
\edbTriple(\constAbbTransRel,\allowbreak\constAbbOwlInvOf,\allowbreak\constAbbTransRelInv)
\}$.
Iteratively applying rules \eqref{eq:ruleedbidb}--\eqref{eq:ruletrans} to $\Inter$,
we obtain the following new derivations in each step, where superscripts indicate the rule used to
produce each fact:
\smallskip

\noindent
\begin{tabular}{@{}llll@{}}
$\aprogram^1(\Inter):$
 & $\idbTriple(\constAbbTransRel, \constAbbOwlInvOf, \constAbbTransRelInv)^{\eqref{eq:ruleedbidb}}$
 & $\idbTriple(\constA,\allowbreak\constAbbTransRel, \constB)^{\eqref{eq:ruleedbidb}}$
 & $\idbTriple(\constB,\allowbreak\constAbbTransRel, \constC)^{\eqref{eq:ruleedbidb}}$
\\
$\aprogram^2(\Inter):$
 & $\idbInv(\constAbbTransRel, \constAbbTransRelInv)^{\eqref{eq:ruleinv}}$
 & $\idbTriple(\constA,\allowbreak\constAbbTransRel, \constC)^{\eqref{eq:ruletrans}}$
\\
$\aprogram^3(\Inter):$
 & $\idbTriple(\constB,\constAbbTransRelInv,\constA)^{\eqref{eq:ruleinvone}}$
 & $\idbTriple(\constC,\constAbbTransRelInv,\constB)^{\eqref{eq:ruleinvone}}$
 & $\idbTriple(\constC,\constAbbTransRelInv,\constA)^{\eqref{eq:ruleinvone}}$
\end{tabular}\medskip

\noindent
No further facts can be inferred. For example,
applying rule \eqref{eq:ruleinvtwo} to $\aprogram^3(\Inter)$ only yields duplicates of
previous inferences. %

\makeatletter{}%
\section{Semi-Naive Evaluation}\label{sec:sne}

Our goal is to compute the materialization $\aprogram^\infty(\Inter)$. For this we use a variant of the well-known technique of \emph{semi-naive evaluation} (SNE)~\cite{Alice} that is based on a more fine-grained notion of derivation step.

In each step of the algorithm, we apply one rule $\arule\in\aprogram$ to the facts derived so far. We do this fairly, so that each rule will be applied arbitrarily often. This differs from standard SNE where all rules are applied in parallel in each step.
We write $\steprule{i}$ for the rule applied in step $i$, and $\Delta^i_p$ for the set of new facts with predicate $p$ derived in step $i$. Note that $\Delta^i_p=\emptyset$ if $p$ is not the head predicate of $\steprule{i}$.
Moreover, for numbers $0\leq i\leq j$, we define the set $\Delta^{[i,j]}_p\defeq \bigcup_{k=i}^j \Delta_p^k$ of all $p$-facts derived between steps $i$ and $j$.
Consider a rule
\begin{align}
\arule= p(\vec{t})\leftarrow e_1(\vec{t_1}),\ldots, e_n(\vec{t_n}), q_1(\vec{s_1}),\ldots,q_m(\vec{s_m})
\label{eq:arule}
\end{align}
where $p,q_1,\ldots,q_m$ are IDB predicates and $e_1,\ldots,e_n$ are EDB predicates. The naive way to apply $\arule$ in step $i+1$ to compute $\Delta^{i+1}_p$ is to evaluate the following ``rule''\footnote{Treating sets of facts like predicates is a common abuse of notation for explaining SNE \cite{Alice}.}
\begin{align}
\text{tmp}_p(\vec{t})\leftarrow e_1(\vec{t_1}),\ldots, e_n(\vec{t_n}), \Delta^{[0,i]}_{q_1}(\vec{s_1}),\ldots,\Delta^{[0,i]}_{q_m}(\vec{s_m})\label{eq:naiverule}
\end{align}
and to set $\Delta^{i+1}_p\defeq\text{tmp}_p\setminus \Delta^{[0,i]}_p$.
However, this would recompute all previous inferences of $\arule$ in each step where $\arule$ is applied. Assume that rule $\arule$ has last been evaluated in step $j<i+1$. We can restrict to evaluating the following rules:
\begin{align}
\begin{split}
\text{tmp}_p(\vec{t})\leftarrow{} & e_1(\vec{t_1}),\ldots, e_n(\vec{t_n}),
   \Delta^{[0,i]}_{q_1}(\vec{s_1}),\ldots,\Delta^{[0,i]}_{q_{\ell-1}}(\vec{s_{\ell-1}}),\\
 & \Delta^{[j,i]}_{q_\ell}(\vec{s_{\ell}}),\Delta^{[0,j-1]}_{q_{\ell+1}}(\vec{s_{\ell+1}}),\ldots, \Delta^{[0,j-1]}_{q_m}(\vec{s_m})
\end{split}\label{eq:snerule}
\end{align}
for all $\ell\in\{1,\ldots,m\}$. With $\text{tmp}_p$ the union of all sets of facts derived from these $m$ rules, we can define $\Delta^{i+1}_p\defeq\text{tmp}_p\setminus \Delta^{[0,i]}_p$ as before.
It is not hard to see that the rules of form \eqref{eq:snerule} consider all combinations of facts that are considered in rule \eqref{eq:naiverule}.
We call this procedure the \emph{one-rule-per-step} variant of SNE. The procedure terminates if all rules in $\aprogram$ have been applied in the last $|\aprogram|$ steps without deriving any new facts.%

\begin{theorem}\label{theo_seminaive}
For every input database instance $\Inter$, and for every fair application strategy of rules, the one-rule-per-step variant of SNE terminates in some step $i$ with the result $\bigcup_{p} \Delta_p^{[0,i]} = \aprogram^\infty(\Inter)$.
\end{theorem}

SNE is still far from avoiding all redundant computations. For example, any strategy of
applying rules \eqref{eq:ruleedbidb}--\eqref{eq:ruletrans} above will
lead to
$\idbTriple(\constB,\constAbbTransRelInv,\constA)$ being derived by rule \eqref{eq:ruleinvone}.
This new inference will be considered
in the next application of the second SNE variant
$\text{tmp}_{\idbTriple}(y,v,x)\leftarrow\Delta_{\idbInv}^{[0,i]}(v,w),\Delta_{\idbTriple}^{[j,i]}(x,w,y)$
of rule \eqref{eq:ruleinvtwo}, leading to the
derivation of $\idbTriple(\constA,\constAbbTransRel,\constB)$. However, this fact must be a duplicate since it is necessary to derive $\idbTriple(\constB,\constAbbTransRelInv,\constA)$ in the first place.
\makeatletter{}%
\section{Column-Oriented Datalog Materialization}

Our variant of SNE provides us with a
high-level materialization procedure. To turn this into an
efficient algorithm, we use a column-based storage layout
described next.

\begin{figure}[tb]
\def\svgwidth{1.4\linewidth}
\resizebox{\linewidth}{!}{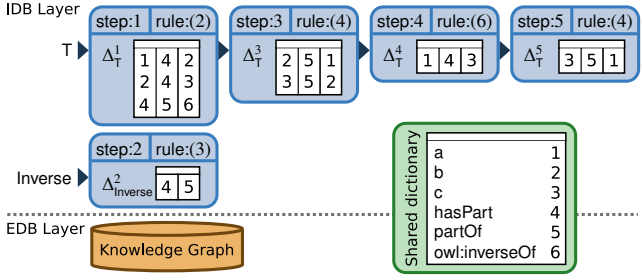}
\caption{Storage Layout for Column-Based Materialization}
\label{fig:storage}
\end{figure}
Our algorithms distinguish the data structures used for storing the 
initial knowledge graph (EDB layer) from those used to 
store derivations (IDB layer), as illustrated in 
Fig.~\ref{fig:storage}. The materialization process accesses the KG by 
asking conjunctive queries to the EDB layer. There are well-known ways 
to implement this efficiently, such as \cite{rdf3x}, and hence we
focus on the IDB layer here.

Our work is inspired by column-based 
databases~\cite{IGNMMK12:MonetDBColumnResearch}, an alternative to traditional
row-based databases for efficiently storing large data volumes.
Their superior performance on analytical queries is compensated for by
lower performance for data updates.
Hence, we structure the IDB layer using a 
column-based layout in a way that avoids the need for 
frequent updates. To achieve this, we store each of the sets of 
inferences $\Delta^i_p$ that are produced during the derivation in a 
separate column-oriented table. The table for $\Delta^i_p$ is created 
when applying $\steprule{i}$ in step $i$ and never modified 
thereafter. We store the data for each rule application (step number, 
rule, and table) in one \emph{block}, and keep a separate list of 
blocks for each IDB predicate. The set of facts derived for one IDB 
predicate $p$ is the union of the contents of all tables in the list 
of blocks for $p$. Figure~\ref{fig:storage} illustrates this scheme, 
and shows the data computed for the running example.

The columnar tables for $\Delta^i_p$ are sorted by extending the order of integer indices used for constants
to tuples of integers in the natural way (lexicographic order of tuples).
Therefore, the first column is fully sorted, the second column is a concatenation of sorted
lists for each interval of tuples that agree on the first component, and so on. Each column is
compressed using run-length encoding (RLE), where maximal sequences of $n$ repeated constants $c$
are represented by pairs $\tuple{a,n}$ \cite{abadi_integrating_2006}.

Our approach enables valuable space savings for in-memory computation.  Ordering
tables improves compression rates, and rules with constants in their heads
(e.g., \eqref{eq:ruletrans}) lead to constant columns, which occupy almost no
memory. Furthermore, columns of EDB relations can be represented by queries
that retrieve their values from the EDB layer, rather than by a copy of these
values. Finally, many inference rules simply ``copy'' data from one
predicate to another, e.g., to define a subclass relationship, so we can often
share column-objects in memory rather than allocating new space.

We also obtain valuable time savings.
Sorting tables means they can be used in \emph{merge joins},
the most efficient type of join, where two sorted relations are compared in a single pass.
This also enables efficient, set-at-a-time duplicate elimination, which we implement by performing outer merge joins between a newly derived result $\textsf{tmp}_p$ and all previously derived tables $\Delta^i_p$.
The use of separate tables for each $\Delta^i_p$ eliminates the cost of
insertions, and at the same time enables efficient \emph{bookkeeping} to record
the derivation step and rule used to produce each inference. Step information is needed to
implement SNE, but the separation of inferences by rule enables further optimizations
(see next section).

There is also an obvious difficulty for using our approach.
To evaluate a SNE rule \eqref{eq:snerule}, we need to
find all answers to the rule's body, viewed as a conjunctive query. This can be achieved by computing the following join:
\begin{equation}
\begin{split}
\big(e_1(\vec{t_1})\;{\bowtie}\ldots{\bowtie}\; e_n(\vec{t_n})\big)\bowtie
   \Delta^{[0,i]}_{q_1}(\vec{s_1})\bowtie\ldots\bowtie\Delta^{[0,i]}_{q_{\ell-1}}(\vec{s_{\ell-1}})\\
 {}\bowtie\Delta^{[j,i]}_{q_\ell}(\vec{s_{\ell}})\bowtie\Delta^{[0,j-1]}_{q_{\ell+1}}(\vec{s_{\ell+1}})\bowtie\ldots\bowtie \Delta^{[0,j-1]}_{q_m}(\vec{s_m})
 \end{split}\label{eq:snejoin}
\end{equation}
The join of the EDB predicates $e_k$ can be computed efficiently by the EDB layer; let $R_{\textsf{EDB}}$ denote the resulting relation. Proceeding from left to right, we now need to compute $R_{\textsf{EDB}}\bowtie\Delta^{[0,i]}_{q_1}(\vec{s_1})$. However, our storage scheme stores the second relation in many blocks, so that we actually must compute
$R_{\textsf{EDB}}\bowtie(\bigcup_{k=0}^i\Delta^{k}_{q_1})(\vec{s_1})$, which could be expensive if there are many non-empty $q_1$ blocks in the range $[0,i]$.

We reduce this cost by performing \emph{on-demand concatenation} of tables: before computing the join, we consolidate $\Delta^{k}_{q_1}$ ($k=0,\ldots,i$) in
a single data structure. This structure is either a hash table or a fully sorted table -- the rule engine decides heuristically to use a hash or a merge join.
In either case, we take advantage of our columnar layout and concatenate only columns needed in the join, often just a single column.
The join performance gained with such a tailor-made data structure justifies the cost of on-demand concatenation. We delete the auxiliary structures after the join.

This approach is used whenever the union of many IDB tables is needed in a join. However, especially the expression $\Delta^{[j,i]}_{q_\ell}$ may often refer to only one (non-empty) block, in which case we can work directly on its data. We use several optimizations that aim to exclude some non-empty blocks from a join so as to make this more likely, as described next. %

\makeatletter{}%
\section{Dynamic Optimization}

Our storage layout is most effective when only a few blocks of fact tables $\Delta_p^i$ must be considered for applying a rule, as this will make on-demand concatenation simpler or completely obsolete.
An important advantage of our approach is that we can exclude individual blocks when applying a rule, based on any information that is available at this time.

We now present three different optimization techniques whose goal is precisely this. In each case, assume that we
have performed $i$ derivation steps and want to apply rule $\arule$ of the form \eqref{eq:arule} in step $i+1$, and that $j<i+1$ was the last step in which $\arule$ has been applied.
We consider each of the $m$ versions of the SNE rule
\eqref{eq:snerule} in separation.
We start by gathering, for each IDB atom $q_k(\vec{s_k})$ in the body of $\arule$, the
relevant range of non-empty tables
$\Delta_{q_k}^o$. We also record which rule
$\steprule{o}$ was used to create this table in step $o$.

\subsection{Mismatching Rules}

An immediate reason for excluding $\Delta_{q_k}^o$ from the join is that the head of
$\steprule{o}$ does not unify with $q_k(\vec{s_k})$.
This occurs when there are distinct constant symbols in the two atoms. In such a case, it is clear that none of the IDB facts in $\Delta_{q_k}^o$ can contribute to matches of $q_k(\vec{s_k})$, so we can safely remove $o$ from the list of blocks considered for this body atom. For example, rule \eqref{eq:ruleinv} can always ignore inferences of rule \eqref{eq:ruletrans}, since the constants $\constTransRel$ and $\constOwlInvOf$
do not match.

We can even apply this optimization if the head of $\steprule{o}$ unifies with the body atom $q_k(\vec{s_k})$, by exploiting the information contained in partial results obtained when computing
the join \eqref{eq:snejoin} from left to right.
Simplifying notation, we can write \eqref{eq:snejoin}
as follows:
\begin{equation}
R_{\textsf{EDB}}\bowtie \Delta_{q_1}^{[l_1,u_1]}\bowtie\ldots\bowtie\Delta_{q_m}^{[l_m,u_m]}
\end{equation}
where $R_{\textsf{EDB}}$ denotes the relation obtained by joining the EDB atoms. We compute this $m$-ary  join by applying $m$ binary joins from left to right.
Thus, the decision about the blocks to include for $\Delta_{q_k}^{[l_k,u_k]}$ only needs to be made when we have already computed the relation $R_k\defeq R_{\textsf{EDB}}\bowtie \Delta_{q_1}^{[l_1,u_1]}\bowtie\ldots\bowtie\Delta_{q_{k-1}}^{[l_{k-1},u_{k-1}]}$. This relation yields all possible instantiations for the variables that occur in
the terms $\vec{t_1},\ldots,\vec{t_n},\vec{s_1},\ldots,\vec{s_{k-1}}$, and we can thus view $R_k$ as a set of possible partial substitutions that may lead to a match of the rule. Using this notation, we obtain the following result.

\begin{theorem}\label{theo_rulemismatch}
If, for all $\sigma\in R_k$, the atom
$q_k(\vec{s_k})\sigma$ does \emph{not} unify with the head of $\steprule{o}$, then the result of \eqref{eq:snejoin} remains the same when replacing
the relation $\Delta_{q_k}^{[l_k,u_k]}$ by
$(\Delta_{q_k}^{[l_k,u_k]}\setminus \Delta_{q_k}^o)$.
\end{theorem}

This turns a static optimization technique into a dynamic, data-driven optimization. While the static approach required a mismatch of rules under all possible instantiations, the dynamic version considers only a subset of those, which is guaranteed to contain all actual matches.
This idea can be applied to other optimizations as well. In any case, implementations must decide
if the cost of checking a potentially large number of partial instantiations in $R_k$ is worth paying
in the light of the potential savings.

\subsection{Redundant Rules}

A rule is \emph{trivially redundant} if its head atom
occurs in its body. Such rules do not need to be applied, as they can only produce duplicate inferences. While trivially redundant rules are unlikely to occur in practice, the combination of two rules frequently has this form. Namely, if the head of $\steprule{o}$ unifies with $q_k(\vec{s_k})$, then we can resolve rule $\arule$ with $\steprule{o}$, i.e., apply backward chaining, to obtain a rule of the form:
\begin{align}
\begin{split}
\arule_o= p(\vec{t})\leftarrow{}& e_1(\vec{t_1}),\ldots, e_n(\vec{t_n}), q_1(\vec{s_1}),\ldots,q_{k-1}(\vec{s_{k-1}}),\\
& \textsf{Body}_{\steprule{o}},q_{k+1}(\vec{s_{k+1}}),\ldots,q_m(\vec{s_m}).
\end{split}
\label{eq:aruleresolved}
\end{align}
where $\textsf{Body}_{\steprule{o}}$ is a variant of the body of $\steprule{o}$ to which a most general unifier has been applied.
If rule $\arule_o$ is trivially redundant, we can again ignore $\Delta_{q_k}^o$. Moreover, we can again turn this into a dynamic optimization method
by using partially computed joins as above.

\begin{theorem}\label{theo_ruleredundant}
If, for all $\sigma\in R_k$, the rule
$\arule_o\sigma$ is trivially redundant, then
the result of \eqref{eq:snejoin} remains the same when replacing the relation $\Delta_{q_k}^{[l_k,u_k]}$ by
$(\Delta_{q_k}^{[l_k,u_k]}\setminus \Delta_{q_k}^o)$.
\end{theorem}

For example, assume we want to apply rule
\eqref{eq:ruleinvtwo} of our initial example,
and $\Delta_{\idbTriple}^o$ was derived
by rule \eqref{eq:ruleinvone}.
Using backward chaining, we obtain
$\arule_o = \idbTriple(y,w,x)\leftarrow \idbInv(v,w),\idbInv(v,w'),\idbTriple(y,w',x)$,
which is not trivially redundant. However, evaluating the first part of the body $\idbInv(v,w),\idbInv(v,w')$ for our initial example data, we obtain just a single substitution
$\sigma=\{v\mapsto\constAbbTransRel,\allowbreak
w\mapsto\constAbbTransRelInv,\allowbreak
w'\mapsto\constAbbTransRelInv\}$.
Now
$\arule_o\sigma = \idbTriple(y,\constAbbTransRelInv,x)\leftarrow \idbInv(\constAbbTransRel,\constAbbTransRelInv),\allowbreak
\idbInv(\constAbbTransRel,\constAbbTransRelInv),\allowbreak
\idbTriple(y,\constAbbTransRelInv,x)$ is trivially redundant. This optimization depends on the data, and cannot be found by considering rules alone.

\subsection{Subsumed Rules}

Many further optimizations can be realized using
our novel storage layout.
As a final example, we present an optimization
that we have not implemented yet, but which we think
is worth mentioning as it is theoretically sound
and may show a promising direction for
future works.
Namely, we consider the case where some of the inferences of
rule $\arule$ were already produced by another
rule since the last application of $\arule$ in step $j$.
We say that rule $\arule_1$ is \emph{subsumed} by rule $\arule_2$ if, for all sets of facts $\Inter$,
$\arule_1(\Inter)\subseteq\arule_2(\Inter)$.
It is easy to compute this, based on the well-known
method of checking subsumption of conjunctive queries
\cite{Alice}.
If this case is detected, $\arule_1$ can be ignored
during materialization, leading to another form of static optimization.
However, this is rare in practice.
A more common case is that one specific way of applying $\arule_1$ is subsumed by $\arule_2$.

Namely, when considering whether to use $\Delta_{q_k}^o$ when applying rule $\arule$,
we can check if the resolved rule $\arule_o$
shown in \eqref{eq:aruleresolved} is
subsumed by a rule $\arule'$ that has already been applied after step $o$. If yes, then $\Delta_{q_k}^o$ can again be ignored.
For example, consider the rules \eqref{eq:ruleedbidb}--\eqref{eq:ruletrans} and an additional rule
\begin{equation}
\idbTransRelDomain(x) \leftarrow \idbTriple(x,\constTransRel,y),
\label{eq:ruledomain}
\end{equation}
which is a typical way to declare the domain of a property. Then we never need to apply rule
\eqref{eq:ruledomain} to inferences of rule
\eqref{eq:ruletrans}, since the combination of these rules
$\idbTransRelDomain(x) \leftarrow \idbTriple(x,\constTransRel,y'),\allowbreak\idbTriple(y',\constTransRel,y)$ is subsumed by rule \eqref{eq:ruledomain}.

One can pre-compute these relationships statically,
resulting in statements of the form ``$\arule_1$ does not need to be applied to inferences produced by $\arule_2$ in step $o$ if $\arule_3$ has already been applied to all facts up until step $o$.'' This information can then be used dynamically during materialization to eliminate further
blocks. The special case $\arule_1=\arule_3$ was illustrated in the example. It is safe for a rule to subsume part of its own application in this way.
\makeatletter{}%
\section{Memoization}

The application of a rule with $m$ IDB body atoms requires the evaluation of $m$ SNE rules of the form
\eqref{eq:snerule}. Most of the joined relations
$\Delta_{q_k}^{[l_k,u_k]}$
range over (almost) all inferences of the respective IDB atom, starting from $l_k=0$. Even if optimizations
can eliminate many blocks in this range, the algorithm may spend considerable resources on computing these optimizations and the remaining on-demand concatenations, which may still be required.
This cost occurs for each application of the rule, even if there were no new inferences for $q_k$ since the last computation.

Therefore, rules with fewer IDB body atoms can be evaluated faster. Especially rules with only one IDB body atom require only a single SNE rule using the limited range of blocks $\Delta_{q_1}^{[j,i]}$.
To make this favorable situation more common,
we can pre-compute the extensions of selected IDB atoms, and then treat these atoms as part of the EDB layer. We say that the pre-computed IDB atom is \emph{memoized}. For example, we could memoize
the atom $\idbTriple(v,\constOwlInvOf,w)$ in \eqref{eq:ruleinv}.
Note that we might memoize an atom without pre-computing all instantiations of its predicate.
A similar approach was used for OWL~RL reasoning by \citeauthor{hybrid} \shortcite{hybrid}, who proved the correctness of this transformation.

SNE is not efficient for selective pre-computations, since it would compute large parts of the materialization.
Goal-directed methods, such as QSQ-R or Magic Sets, focus on inferences needed to answer a given query
and hence are more suitable \cite{Alice}.
We found QSQ-R to perform best in our setting.

Which IDB atoms should be memoized?
For specific inferencing tasks, this choice is often fixed.
For example, it is very common to pre-compute the sub-property hierarchy.
We cannot rely on such prior domain knowledge for general Datalog, and we
therefore apply a heuristic: we attempt pre-computation for all 
most general body atoms with QSQ-R, but set a timeout (default 1 sec).
Memoization is only performed for atoms where pre-computation completes before this time.
This turns out to be highly effective in some cases. %

\makeatletter{}%
\section{Evaluation}
\label{sec:evaluation}

\begin{table}[tb]
\footnotesize
\begin{tabular}{l | r | r | r |r |r}
 & \textbf{\#triples} & \bf \toolname & \multicolumn{3}{c}{\textbf{Rule sets}}\\
\textbf{Dataset} & \textbf{(EDB facts)} & \textbf{DB size} & \textbf{L}& \textbf{U}& \textbf{LE}\\
\hline
\tableTopSpace
LUBM1K   & 133M & 5.5GB & 170 & 202 & 182 \\
LUBM5K   & 691M &  28GB & "  & " & "\\
DBpedia  & 112M & 4.8GB & 9396 & --- & --- \\
Claros   &  19M & 980MB & 2689 & 3229 & 2749\\
Claros-S & 500K &  41MB & "  & " & "\\
\end{tabular}

\caption{Statistics for Datasets and Rule Sets Used}
\label{tab:info}
\end{table}

In this section, we evaluate our approach based on a prototype implementation called
\emph{\toolname}. %
As our main goal is to support KG materialization under limited
resources, we perform all evaluations on a laptop computer.
Our source code and a short tutorial is found at \url{https://github.com/jrbn/vlog}.

\fakeparagraph{Experimental Setup} The computer used in all experiments is a
Macbook Pro with a 2.2GHz Intel Core i7 processor, 512GB SDD, and 16GB RAM
running on MacOS Yosemite OS v10.10.5. All software (ours and competitors) was
compiled from C++ sources using Apple CLang/LLVM v6.1.0.

We used largely the same data that was also used to evaluate RDFox \cite{rdfox}. Datasets
and Datalog programs are available online.\footnote{\url{http://www.cs.ox.ac.uk/isg/tools/RDFox/2014/AAAI/}}
The datasets we used are the cultural-heritage ontology Claros \cite{rdfox}, the
DBpedia KG extracted from Wikipedia \cite{dbpedia09}, and two differently sized graphs generated with the LUBM benchmark~\cite{lubm}. In addition, we created
a random sample of Claros that we call Claros-S.
Statistics on these datasets are given in Table~\ref{tab:info}.

All of these datasets come with OWL ontologies that can be used for inferencing.
\citeauthor{rdfox} used a custom translation of these ontologies into Datalog.
There are several types of rule sets: ``L'' denotes the custom translation of the
original ontology; ``U'' is an (upper) approximation of OWL ontologies that cannot
be fully captured in Datalog; ``LE'' is an extension of the ``L'' version with
additional rules to make inferencing harder. All of these rules operate on a
Datalog translation of the input graph, e.g., a triple
$\tuple{\textsf{entity:5593},\allowbreak\textsf{rdf:type},\allowbreak\textsf{a3:Image}}$ might be represented by a fact $\textsf{a3:Image}(\textsf{entity:5593})$.
We added rules to translate EDB triples to IDB atoms.
The W3C standard also defines another set of derivation rules for OWL~RL that can work directly on triples \cite{owl2-profiles}. We use ``O'' to refer to 66 of those rules, where we omitted
the rules for datatypes and equality reasoning \cite[Tables 4 and 8]{owl2-profiles}.

\toolname combines an on-disk EDB layer with an in-memory columnar IDB 
layer to achieve a good memory/runtime balance on limited hardware. 
The specifically developed on-disk database uses six permutation 
indexes, following standard practice in the field~\cite{rdf3x}. No 
other tool is specifically optimized for our setting, but the leading 
in-memory system RDFox is most similar, and we therefore use it for 
comparison.
As our current prototype does not use parallelism, we compared it to 
the sequential version of the original version of RDFox \cite{rdfox}. 
We recompiled it with the ``release'' configuration and the sequential 
storage variant.
Later RDFox versions perform equality reasoning, which would lead to 
some input data being interpreted differently 
\shortcite{MNPH15:rdfoxIncrementalEquality,MNPH15:rdfoxEqualityRewriting}.  We were unable to deactivate this 
feature, and hence did not use these versions. If not stated 
otherwise, \toolname was always used with dynamic optimizations 
activated but without memoization.

\newcommand{\oom}{\textsf{oom}\xspace}
\newcommand{\timeout}{\textsf{tout}\xspace}
\begin{table}[tb]
    \footnotesize
\mbox{}\hfill\begin{tabular}{@{}l | r | r | r | r| r@{}}
        \bf Data/Rules &
            \multicolumn{2}{c|}{\textbf{RDFox (seq)}} &
            \multicolumn{3}{c}{\textbf{\toolname}}
        \\
         & time & mem & time & mem & IDBs\\
        \hline\tableTopSpace
        LUBM1K/L & 82 & 11884 & 38 & 2198 & 172M \\
        LUBM1K/U & 148 & 14593 & 80 & 2418 & 197M\\
        LUBM1K/LE & \oom & \oom & 2175 & 9818 & 322M \\
        LUBM5K/L & \oom & \oom & 196 & 8280 & 815M\\
        LUBM5K/U & \oom & \oom & 434 & 7997 & 994M\\
        LUBM5K/LE & \oom & \oom & \timeout & \timeout & ---\\
        DBpedia/L & 177 & 7917 & 91 & 532 & 33M\\
        Claros/L & 2418 & 5696 & 644 & 2406 & 89M\\
        Claros/LE & \oom & \oom & \timeout & \timeout & ---\\
        Claros-S/LE & 8.5 & 271 & 2.5 & 127 & 3.7M\\
    \end{tabular}\hfill\mbox{}
    \caption{Materialization Time (sec) and Peak Memory (MB)}
    \label{tab:overviewperf}
\end{table}
\fakeparagraph{Runtime and Memory Usage} Table~\ref{tab:overviewperf} reports the
runtime and memory usage for materialization on our test data, and the total number of inferences
computed by \toolname.
Not all operations could be completed on our hardware:
\oom denotes an out-of-memory error, while \timeout denotes a timeout after 3h.
Memory denotes the peak RAM usage as measured using OS APIs.

The number of IDB facts inferred by \toolname
is based on a strict separation of IDB and EDB predicates, using rules like \eqref{eq:ruleedbidb}
to import facts used in rules. This is different from
the figure reported for RDFox, which corresponds to unique triples (inferred or given). We have compared
the output of both tools to ensure correctness.

RDFox has been shown to achieve excellent speedups using multiple CPUs, so our
sequential runtime measurements are not RDFox's best
performance but a baseline for fast in-memory computation in a single thread.
Memory usage can be compared more directly, since
the parallel version of RDFox uses only slightly more memory \cite{rdfox}.
As we can see, \toolname requires only 6\%--46\% of the working memory used by RDFox.
As we keep EDB data on disk, the comparison with a pure in-memory system like RDFox should take the on-disk file sizes into account (Table~\ref{tab:info}); even when we add these, \toolname uses less memory in all cases where
RDFox terminates.
In spite of these memory savings, \toolname shows
comparable runtimes, even when considering an (at most linear) speedup when parallelizing RDFox.

\begin{table}[tb]%
\footnotesize%
\mbox{}\hfill\begin{tabular}{l | r | r | r | r}
\bf Data/Rules & \bf MR+RR  & \bf MR & \bf RR & \bf No opt. \\
\hline\tableTopSpace
LUBM1K/L & 38 & 39 & 38 & 40 \\
LUBM5K/L & 196 & 197 & 202 & 206 \\
DBpedia/L & 91 & 92 & 93 & 88 \\
Claros/L & 644 & 3130 & 684 & 3169 \\
\end{tabular}\hfill\mbox{}
\caption{Impact of Dynamic Optimizations (times in sec)}
\label{tab:bpstrat}
\end{table}
\fakeparagraph{Dynamic Optimization} Our prototype supports the optimizations
``Mismatching Rules'' (MR) and ``Redundant Rules'' (RR)
discussed earlier.
Table~\ref{tab:bpstrat} shows the runtimes obtained
by enabling both, one, or none of them.

Both MR and RR have little effect on LUBM and DBpedia. We attribute this to the rather ``shallow'' rules used in both cases.
In constrast, both optimizations are very effective
on Claros, reducing runtime by a factor of almost five. This is because SNE leads to some expensive joins that produce only duplicates and that the optimizations can avoid.

\begin{table}[tb]
\footnotesize
\begin{tabular}{r | r | r | r | r | r}
\bf Data/Rules & \bf No Mem. & \multicolumn{4}{c}{\bf Memoization}\\
& \bf $t_{\text{total}}$ & \#atoms & \bf $t_{\text{Mem}}$ & \bf $t_{\text{Mat}}$ & \bf $t_{\text{total}}$ \\
\hline\tableTopSpace
LUBM1K/L & 38 & 39 & 1.4 & 40.4 & 41.5\\
LUBM1K/O & 1514 & 41 & 6.5 & 230 & 236.5 \\
\end{tabular}
\caption{Impact of Memoization (times in sec)}
\label{tab:memo}
\end{table}
\fakeparagraph{Memoization} To evaluate the impact of memoization, we
materialized LUBM1K with and without this feature,
using the L and O rules.
Table~\ref{tab:memo} shows total runtimes
with and without memoization, the number of
IDB atoms memoized, and the time used to compute their
memoization.

For the L rules, memoization has no effect on
materialization runtime despite the fact that 39 IDB atoms were memoized.
For the O rules, in contrast, memoization decreases materialization runtime by a factor of six, at an initial cost of 6.5 seconds. We conclude that this procedure is indeed beneficial, but only if we use the standard OWL~RL rules. Indeed, rules such as \eqref{eq:ruleinvone}, which we used to motivate memoization, do not occur in the L rules.
In a sense, the construction of L rules internalizes
certain EDB facts and thus pre-computes their effect before materialization. %

\makeatletter{}%
\section{Discussion and Conclusions}
\label{sec:conclusions}
We have introduced a new column-oriented approach to perform Datalog in-memory materialization over large KGs. Our goal was to perform this task in an efficient manner, minimizing memory consumption and CPU power. Our evaluation indicates that it is a viable alternative to existing Datalog engines, leading to competitive runtimes at a significantly reduced memory consumption.

Our evaluation has also highlighted some challenges to address in 
future work. First, we observed that the execution of large joins can 
become problematic when many tables must be scanned for removing 
duplicates. This was the primary reason why the computation did not 
finish in time on some large datasets. Second, our implementation does 
not currently exploit multiple processors, and it will be interesting 
to see to how techniques of intra/inter query parallelism can be 
applied in this setting. Third, we plan to study mechanisms for 
efficiently merging inferences back into the input KG, which is not 
part of Datalog but useful in practice. Finally, we would also like to 
continue extending our dynamic optimizations to more complex cases, 
and to develop further optimizations that take advantage of our design.

Many further continuations of this research come to mind.
To the best of our knowledge, this is the first work
to exploit a column-based approach for Datalog inferencing,
and it does indeed seem as if the research on
large-scale in-memory Datalog computation has only just begun. %

{
\fakeparagraph{Acknowledgments} This work was partially funded by COMMIT, the
NWO VENI project 639.021.335, and the DFG in Emmy Noether grant KR~4381/1-1 and in CRC~912 \emph{HAEC} within the \emph{cfAED} Cluster of Excellence.
}

\newpage
\bibliographystyle{aaai}

\begin{report}
\makeatletter{}%
\newpage
\appendix

\section{Appendix: Proofs}

\subsection{Proof of Theorem~\ref{theo_seminaive}}

We first observe that the naive approach \eqref{eq:naiverule} terminates
and leads to a unique least model $\aprogram^\infty(\Inter)$. Recall that
the latter was defined by applying all rules in parallel in each step.
Now consider an arbitrary, fair sequence of
individual applications of rules $\steprule{1},\steprule{2},\ldots$, each applied 
naively as in \eqref{eq:naiverule}.
Let $\hat{\aprogram}^\ell(\Inter)$ denote the set of all facts derived in this way up 
until step $\ell$.
Clearly, the rule-by-rule inference is sound, i.e.,
$\hat{\aprogram}^\ell(\Inter)\subseteq\aprogram^\infty(\Inter)$ for all
derivation steps $\ell$.
It remains to show that it is also complete in the sense that
$\hat{\aprogram}^\ell(\Inter)\supseteq\aprogram^\infty(\Inter)$ for some $\ell$.
Since we apply rules fairly, there is a sequence of derivation
step indices $i_1< i_2 < i_3 < \ldots $ such that every rule has been applied
in each interval of the form $i_k<i_{k+1}$. Formally, for every $i_k$ in the
sequence, and for every rule $\arule$, there is $j\in\{i_k+1,\ldots,i_{k+1}\}$ 
such that $\arule=\steprule{j}$.
It follows that $\hat{\aprogram}^{i_{k+1}}(\Inter)\supseteq\aprogram(\hat{\aprogram}^{i_{k}}(\Inter))$
(in words: the sequential application of rules  derives at least the inferences that a parallel
application of rules would derive).
Therefore, by a simple induction, $\aprogram^\ell(\Inter)\subseteq\hat{\aprogram}^{i_{\ell}}(\Inter)$
for every $\ell\geq 0$.
Since $\aprogram^\infty(\Inter)=\aprogram^m(\Inter)$ for some finite $m$ \cite{Alice}, we have
$\aprogram^\infty(\Inter)\subseteq\hat{\aprogram}^{i_{m}}(\Inter)$. Together with soundness,
this implies that $\aprogram^\infty(\Inter)=\hat{\aprogram}^{i_{m}}(\Inter)$, as required.

Now to show that the semi-naive application strategy based on rules of the form
\eqref{eq:snerule} is also sound, we merely need to show that it produces the same
inferences as the naive rule-by-rule application would produce (based on the same, fair
sequence of rules). Let $\Delta^i_p$ refer to the facts derived for $p$ in step
$i$ using the semi-naive procedure, and let $\hat{\Delta}^i_p$ denote the set of facts
produced for $p$ in step $i$ by the naive procedure.
We show by induction that $\Delta^{[0,i]}_p=\hat{\Delta}^{[0,i]}_p$ holds for 
ever predicate $p$ and every step $i$.

The induction base is trivial, since  $\Delta^{0}_p=\hat{\Delta}^{0}_p=\emptyset$.
For the induction step, assume that the claim holds for all $k\leq i$. Let
$\arule$ of form \eqref{eq:arule} be the rule applied in step $i+1$, and assume
that $\arule$ was last applied in step $j$ (set to $-1$ if it was never applied).
$\Delta^{i+1}_p$ is computed by evaluating \eqref{eq:snerule} for every $\ell\in\{1,\ldots, m\}$,
while $\hat{\Delta}^{0}_p$ is obtained by evaluating \eqref{eq:naiverule}.

For every inference $\text{tmp}_p(\vec{c})$ obtained from \eqref{eq:naiverule}, there is
a ground substitution $\sigma$ such that the rule
\begin{align*}
\text{tmp}_p(\vec{t})\sigma\leftarrow e_1(\vec{t_1})\sigma,\ldots, e_n(\vec{t_n})\sigma, \Delta^{[0,i]}_{q_1}(\vec{s_1})\sigma,\ldots,\Delta^{[0,i]}_{q_m}(\vec{s_m})\sigma
\end{align*}
is applicable and $\vec{t}\sigma=\vec{c}$. Being applicable here means that
$\vec{s_a}\sigma\in\Delta^{[0,i]}_{q_a}$ for every $a\in\{1,\ldots,m\}$ (and likewise
for expressions $e_b(\vec{t_b})\sigma$).
Now whenever $\vec{s_a}\sigma\in\Delta^{[0,i]}_{q_a}$, there is an index $w_a\in\{0,\ldots,i\}$
such that $\vec{s_a}\sigma\in\Delta^{w_a}_{q_a}$.

Given an inference $\text{tmp}_p(\vec{c})$ and ground substitution 
$\sigma$ as above, $\text{tmp}_p(\vec{c})$ is also inferred by
a rule of the form \eqref{eq:snerule}. Indeed, let $\ell$ be the largest
index from the range $\{1,\ldots,m\}$ such that $w_\ell\geq j$.
Then the following ground instantiation of \eqref{eq:snerule}
is applicable:
\begin{align*}
\begin{split}
\text{tmp}_p(\vec{t})\sigma\leftarrow{} & e_1(\vec{t_1})\sigma,\ldots, e_n(\vec{t_n})\sigma,\\
 & \Delta^{[0,i]}_{q_1}(\vec{s_1})\sigma,\ldots,\Delta^{[0,i]}_{q_{\ell-1}}(\vec{s_{\ell-1}})\sigma,\\
 & \Delta^{[j,i]}_{q_\ell}(\vec{s_{\ell}})\sigma,\Delta^{[0,j-1]}_{q_{\ell+1}}(\vec{s_{\ell+1}})\sigma,\ldots, \Delta^{[0,j-1]}_{q_m}(\vec{s_m})\sigma.
\end{split}
\end{align*}
This follows from the induction hypothesis and the definition of $\ell$. Note that the case
where $w_a<j$ for all $a\in\{1,\ldots,m\}$ can be disregarded, since it follows by the induction
hypothesis that such inferences have already been produced when applying
rule $\arule$ in step $j$. This completes the induction and the proof.\qed

\subsection{Proof of Theorem~\ref{theo_rulemismatch}}

This claim is immediate from the definitions.
In detail, consider $R_k$ and $\steprule{o}$ as in the claim of the theorem.
Moreover, let $R_m$ be the set of all complete rule body matches
that could be computed without taking any optimization into account.
Clearly, $R_k\bowtie R_m\subseteq R_m$, i.e., $R_m$ contains only
tuples compatible with $R_k$. By the assumption in the theorem,
for all $\sigma\in R_k$, the atom
$q_k(\vec{s_k})\sigma$ does \emph{not} unify with the head of $\steprule{o}$.
Therefore, $\Delta_{q_k}^o$ does not contain any fact that is compatible
with $R_k$, i.e., $R_k\bowtie \Delta_{q_k}^o=\emptyset$ (where the join here is
meant to join the positions in accordance with the terms used in $q_k(\vec{s_k})$).
This implies that $R_m\bowtie \Delta_{q_k}^o=\emptyset$, and thus $\Delta_{q_k}^o$
does not need to be considered for finding matches of $q_k(\vec{s_k})$ when
computing $R_m$. \qed

\subsection{Proof of Theorem~\ref{theo_ruleredundant}}

The claim is again rather immediate, but we spell it out in detail for completeness. Assume we apply a rule $\arule$ of the form
\eqref{eq:arule} in step $i+1$, after it was last applied in step $j$. We use similar notation
for (partial) joins as in the proof of Theorem~\ref{theo_rulemismatch} in the previous section.
In addition, let $\steprule{o}$ be of the following form:
\begin{align*}
\steprule{o}= q_k(\vec{t'})\leftarrow e'_1(\vec{t'_1}),\ldots, e'_{n'}(\vec{t'_{n'}}), q'_1(\vec{s'_1}),\ldots,q'_{m'}(\vec{s'_{m'}}).
\end{align*}
As shown in Theorem~\ref{theo_seminaive},
$\Delta^{[0,o]}_{q_k}$ is the same set of facts that would be produced by evaluating a naive version
of $\steprule{o}$ in step $o$, i.e., by using a computation of the form
\begin{align*}
\text{tmp}_{q_k}(\vec{t'})\leftarrow e'_1(\vec{t'_1}),\ldots, e'_{n'}(\vec{t'_{n'}}), \Delta^{[0,o]}_{q'_1}(\vec{s'_1}),\ldots,\Delta^{[0,o]}_{q'_{m'}}(\vec{s'_{m'}}).
\end{align*}
Note that $\Delta_{q_k}^o\subseteq \text{tmp}_{q_k}(\vec{t'})$.
Let $R'$ denote the result of the following join
\[
e'_1(\vec{t'_1})\bowtie\ldots\bowtie e'_{n'}(\vec{t'_{n'}})\bowtie \Delta^{[0,o]}_{q'_1}(\vec{s'_1})\bowtie\ldots\bowtie\Delta^{[0,o]}_{q'_{m'}}(\vec{s'_{m'}}).
\]
We can again consider the element of $R'$ as substitutions over the variables of $\steprule{o}$, where
we assume without loss of generality that $\steprule{o}$ shares no variables with $\arule$.

Now consider the situation as in the claim where we apply a particular semi-naive rule of the form
\eqref{eq:snerule} and have partially evaluated the rule body until $R_k$.
Consider any $\sigma\in R_k\bowtie R'$ (where the join identifies positions/variables as necessary to unify the atoms $q_k(\vec{t'})$ and $q_k(\vec{s_1})$).
By the definition of redundancy, $\steprule{o}$ contains an atom $q'_a(\vec{s'_a})$
such that $p(\vec{t})\sigma = q'_a(\vec{s'_a})\sigma$ (in particular $p=q'_a$).
As $R'$ assigns values to all variables in $\steprule{o}$, we find that
$\vec{s'_a}\sigma=\vec{t}\sigma$ is a list of ground terms. By definition of $R'$,
$\vec{t}\sigma\in\Delta^{[0,o]}_{q'_a}=\Delta^{[0,o]}_{p}$.
Since $\Delta^{i+1}_p=\text{tmp}_{p}\setminus\Delta^{[0,i]}_p$ and $\Delta^{[0,o]}_{p}\subseteq\Delta^{[0,i]}_p$,
we get $\vec{t}\sigma\notin\Delta^{i+1}_p$.
Therefore, applying rule $\arule$ with any substitution that extends $\sigma$ in step $i+1$ is redundant.
Since the argument holds for all assignments in $R_k\bowtie R'$, and since the projection to of $R_k\bowtie R'$ to
variables in $\arule$ is a superset of $R_k\bowtie \Delta_{q_k}^o$, we find that all tuples from $\Delta_{q_k}^o$
can be ignored when applying $\arule$.\qed

\end{report}

\end{document}